\documentclass[oneside,final,onecolumn]{elsart}
\usepackage{amsmath}

\input{tcilatex}

\begin{document}

\begin{frontmatter}

\title{Practical measurement of joint weak values and their connection to
the annihilation operator}

\author[j]{J.S. Lundeen\corauthref{cor}},
\corauth[cor]{Corresponding author.}
\ead{lundeen@physics.utoronto.ca}
\author[k]{K.J. Resch}
\address[j]{Department of Physics, University of Toronto, 60 St. George Street,
Toronto ON M5S 1A7, Canada}
\address[k]{Institut f\"{u}r Experimentalphysik, Universit\"{a}t Wien,
Boltzmanngasse 5, A-1090 Vienna, Austria}

\begin{abstract}
Weak measurements are a new tool for characterizing post-selected quantum
systems during their evolution. Weak measurement was originally formulated
in terms of von Neumann interactions which are practically available for
only the simplest single-particle observables. In the present work, we
extend and greatly simplify a recent, experimentally feasible, reformulation
of weak measurement for multiparticle observables 
[Resch and Steinberg (2004, Phys. Rev. Lett., 92, 130402)]. We also show that the resulting
``joint weak values'' take on a particularly elegant form when expressed in
terms of annihilation operators.
\end{abstract}

\begin{keyword}
Weak measurement \sep Entanglement \sep Post-selection
\sep Annihilation Operator \PACS 03.65.Ta \sep 02.50.-r \sep 06.30.-k 
\sep 03.67.-a
\end{keyword}

\end{frontmatter}Weak measurement was originally proposed by Aharanov,
Albert and Vaidman (AAV) \cite{AAV} as an extension to the standard von
Neumann (``strong'') model of measurement. \ \ A weak measurement can be
performed by sufficiently reducing the coupling between the measuring device
and the measured system. \ \ In this case, the pointer of the measuring
device begins in a state with enough position uncertainty that any shift
induced by the weak coupling is insufficient to distinguish between the
eigenvalues of the observable in a single trial. \ \ While at first glance
it may seem strange to desire a measurement technique that gives less
information than the standard one, recall that the entanglement generated
between the quantum system and measurement pointer is responsible for
collapse of the wavefunction. \ Furthermore, if multiple trials are
performed on an identically-prepared ensemble of systems, one can measure
the average shift of the pointer to any precision -- this average shift is
called the weak value. \ A surprising characteristic of weak values is that
they need not lie within the eigenvalue spectrum of the observable and can
even be complex \cite{peres,leggett,AV}. \ On the other hand, an advantage
of weak measurements is that they do not disturb the measured system nor any
other simultaneous weak measurements or subsequent strong measurements, even
in the case of non-commuting observables. \ This makes weak measurements
ideal for examining the properties and evolution of systems before
post-selection and might enable the study of new types of observables. \
Weak measurements have been used to simplify the calculation of optical
networks in the presence of polarization-mode dispersion \cite{gisin},
applied to slow- and fast-light effects in birefringent photonic crystals %
\cite{chiao}, and bring a new, unifying perspective to the tunneling-time
controversy \cite{aeweak,aetree}. \ Hardy's Paradox, introduced in Ref. \cite%
{Hardy}, was analyzed in terms of weak values in \cite{Hardyweak}. \ In Ref. %
\cite{wisemanQED}, weak values were used to physically explain the results
of the cavity QED experiment described in \cite{orozco}. \ The opposing
views expressed in Refs. \cite{scully,walls} on the role of which-path
information and the Heisenberg uncertainty principle in the double-slit
experiment are reconciled with the use of weak values in \cite{wisemanweak}.
\ Weak measurement can be considered the best estimate of an observable in a
pre and post-selected system \cite{estimator}. \ 

The von Neumann interaction was originally used to model standard quantum
measurement by mathematically describing the coupling between the measured
system and the measurement pointer \cite{vn}. \ The interaction couples an
observable $\hat{A}$ of the quantum system to the momentum $\hat{P}$ of the
pointer,

\begin{equation}
\mathcal{H}=g\hat{A}\hat{P},  \label{Hamiltonian}
\end{equation}%
where $g$ is the coupling constant which is assumed to be real to keep $%
\mathcal{H}$ Hermitian. \ \ Since\ $\hat{A}$ and $\hat{P}$ act in different
Hilbert spaces we can safely assume they commute. \ This interaction would
be difficult to implement were it not for the fact that typically the
measured system itself is used as part of the measurement device. \ When
measuring $\hat{A}$ of a particle an independent degree of freedom of the
particle can be used as the pointer. \ For example, a birefringent crystal
can be oriented so that it will displace the position of photon by an amount
that depends on the photon's polarization \cite{hulet}. \ Here, $\hat{A}$ is
the polarization observable and the pointer is the position of the photon. \
Another example is the Stern-Gerlach apparatus, where $\hat{A}$ is the spin
of the particle and the pointer is the momentum of the particle.\ If such a
measurement strategy were not availabe, one would require a strong
controllable interaction between the quantum system and a separate pointer
system. \ This is typically far too technically difficult to implement.

In modern quantum mechanics, we are increasingly interested in a different
class of observables than in the above example, in which only a single
particle is involved. \ Often, one would like to measure correlations
between observables of distinct particles, like $\hat{S}_{1}\hat{S}_{2},$
the spin of particle one times the spin of particle two. \ Moreover, any
experiment that utilizes or directly measures properties of entanglement is
based on such observables and so, much of quantum information and quantum
optics deal with these composite or joint observables. \ The exciting
results and complex, rich range of features discovered by studies of
entanglement suggests that weak measurement of joint observables should also
produce valuable and interesting results. \ In fact, there already exist a
few theoretical ideas for weak measurements that center around joint
observables, such as Hardy's Paradox \cite{Hardyweak}, nonlocality of a
single particle \cite{aetree}, and extensions of the Quantum Box Problem %
\cite{3boxexp,aewheeler}. \ We call the weak value of a joint observable the
``joint weak value.'' If the composite observable is a product of $N$ single
particle observables then the weak value is called the ``$N$th-order joint
weak value''.

Joint observables are extremely difficult to measure directly with either
strong or weak types of measurement. \ The difficulty lies in the fact that
the necessary von Neumann interaction couples two separate observables, and
hence particles, to a single pointer. One, therefore, can no longer use the
extra degree of freedom on one of the particles as the pointer and\ so, one
requires multiparticle interactions. \ An approach using multiparticle
interactions was outlined in a proposal for a weak measurement experiment
with ions but so far there have been no experimental weak measurements of
joint observables \cite{molmer}. \ On the other hand, experimental strong
measurements of joint observables are feasible and even commonplace. \ This
is made possible by employing a different measurement strategy. \ Instead of
measuring the joint observable directly, each single\ particle observable is
measured simultaneously but separately. \ For example, instead of measuring $%
\hat{S}_{1}\hat{S}_{2}$ directly we can measure $\hat{S}_{1}$ and $\hat{S}%
_{2}$ separately and then multiply the results trial by trial. \ If one
wants to strongly measure the joint observable $\hat{A}_{1}\hat{A}_{2}...%
\hat{A}_{N}=\hat{M}$, instead of using the multiparticle von Neumann
Hamiltonian $\mathcal{H}=g\hat{M}\hat{P}$, the general strategy is to
simultaneously apply $N$ standard single-particle von Neumann interaction
Hamiltonians,

\begin{eqnarray}
\mathcal{H} &=&g_{1}\hat{A}_{1}\hat{P}_{1}+g_{2}\hat{A}_{2}\hat{P}_{2}+... \\
&=&\sum_{j=1}^{N}g_{j}\hat{A}_{j}\hat{P}_{j}.  \label{nvn}
\end{eqnarray}%
Given that we can already perform each of the single-particle Hamiltonians,
it is straightforward to implement the total Hamiltonian. \ This strategy
allows one to make projective measurements of $\hat{M}$ which is all that is
required to measure the expectation value of $\hat{M}$, \ 
\begin{equation}
\left\langle \hat{M}\right\rangle =\left\langle \hat{A}_{1}\hat{A}_{2}...%
\hat{A}_{N}\right\rangle \propto \left\langle \hat{X}_{1}\hat{X}_{2}...\hat{X%
}_{N}\right\rangle ,  \label{strong}
\end{equation}%
where $\hat{X}_{i}$ is the position operator of the pointer and provided all 
$\hat{A}_{i}$ commute. \ In other words, the expectation value of $\hat{M}$
is related to the correlation between the positions of all $N$ pointers.

In two earlier works, an analogous strategy was applied to weak measurements %
\cite{twoweak,nweak}. \ The Hamiltonian in Eq. (\ref{nvn}) is utilized in
the weak regime to create correlations in the deflections of the $N$
pointers proportional to the weak value. \ Specifically, the $N$th-order
joint weak value was related to two correlations between all $N$ pointer
deflections and a complicated combination of lower-order joint weak values.
\ In this work, we show that the $N$th-order joint weak value takes on an
elegant and simple form closely related to the strong measurement formula in
Eq. (\ref{strong}) when expressed entirely in terms of $N$-pointer
correlations. \ This new and simplified form lends itself to a new way of
thinking about single and joint weak measurements in terms of expectation
values of products of annihilation operators.

We begin by deriving AAV's formula for the weak value of a single particle
observable. \ AAV based weak measurement on the weak limit of the standard
approach to measurement. \ Specifically, they use the von Neumann
interaction in Eq. (\ref{Hamiltonian}),\ which we assume to be constant over
some interaction time $t$. \ The measurement pointer is initially in a
Gaussian wavefunction centered at zero, \ 
\begin{equation}
\ \left\langle x|\phi \right\rangle =\phi (x)=\left( \frac{1}{\sqrt{2\pi }%
\sigma }\right) ^{\frac{1}{2}}\exp \left( -\frac{x^{2}}{4\sigma ^{2}}\right)
,  \label{pointer}
\end{equation}%
where $\sigma $ is the rms width of $\left| \phi (x)\right| ^{2}$.\ \ In
most experiments, quantum mechanical systems are initially prepared in a
known initial state $\left| I\right\rangle $. Since this preparation usually
involves measuring an ensemble of systems and selecting the subensemble with
the correct outcome, this is called pre-selection. \ For a strong
measurement, the von Neumann interaction with a pre-selected system state
shifts the mean position of the pointer $\left\langle \hat{X}\right\rangle $
by $gt\left\langle I\right| \hat{A}\left| I\right\rangle $ and leaves $%
\left\langle \hat{P}\right\rangle $ unchanged. \ AAV considered the case
where we further restrict ourselves to the subensemble of system states that
are found to be in $\left| F\right\rangle $ after the measurement, a
procedure called post-selection. \ A weak measurement performed between the
pre and post-selection can result in very different expectation values than
in strong measurements, as we will see. \ 

After the pointer weakly interacts with the initial sytem-pointer state $%
\left| \psi (0)\right\rangle =\left| I\right\rangle \left| \phi
\right\rangle $ the state evolves to

\begin{eqnarray}
\left| \psi (t)\right\rangle &=&\exp \left( \frac{-i\mathcal{H}t}{\hbar }%
\right) \left| I\right\rangle \left| \phi \right\rangle =\left( 1-\frac{i%
\mathcal{H}t}{\hbar }-...\right) \left| I\right\rangle \left| \phi
\right\rangle  \label{expansion} \\
&=&\left| I\right\rangle \left| \phi \right\rangle -\frac{igt}{\hbar }\hat{A}%
\left| I\right\rangle \hat{P}\left| \phi \right\rangle -...
\end{eqnarray}%
We project out the part of the state that is post-selected in state $\left|
F\right\rangle ,$ 
\begin{equation}
\left\langle F\right| \exp \left( \frac{-i\mathcal{H}t}{\hbar }\right)
\left| I\right\rangle \left| \phi \right\rangle =\left\langle
F|I\right\rangle \left| \phi \right\rangle -\frac{igt}{\hbar }\left\langle
F\right| \hat{A}\left| I\right\rangle \hat{P}\left| \phi \right\rangle -...
\end{equation}%
This leaves the state of pointer after the interaction and post-selection. \
In the limit of an ideal weak measurement, $gt\rightarrow 0,$ $\left|
\left\langle F|I\right\rangle \right| ^{2}=$Prob$_{\text{success}}$ is the
probability the post-selection succeeds \cite{aeweak}. \ If we renormalize
the state and then truncate the amplitude of each term to lowest order in $%
gt $ we get 
\begin{equation}
\left| \phi _{\text{$fi$}}\right\rangle =\left| \phi \right\rangle -\frac{igt%
}{\hbar }\frac{\left\langle F\right| \hat{A}\left| I\right\rangle }{%
\left\langle F|I\right\rangle }\hat{P}\left| \phi \right\rangle -...,
\label{post select pointer}
\end{equation}%
which is just equivalent to dividing by $\left\langle F|I\right\rangle =%
\sqrt{\text{Prob}_{\text{success}}}$. \ The subscript $fi$, corresponding to
final state $\left| F\right\rangle $ and intial state $\left| I\right\rangle 
$, labels the final pointer state, with which we can now calculate the
expectation value of $\hat{X}$ of the pointer. \ The terms which contain an
expectation value of an odd number of operators go to zero since the pointer
is initially an even function about zero. \ To first order in $gt$, the
remaining terms give us \ 
\begin{eqnarray}
\left\langle \hat{X}\right\rangle _{\text{$fi$}} &=&\left\langle \phi _{%
\text{$fi$}}\right| \hat{X}\left| \phi _{\text{$fi$}}\right\rangle =\frac{%
-igt}{\hbar }\func{Re}\left( \frac{\left\langle F\right| \hat{A}\left|
I\right\rangle }{\left\langle F|I\right\rangle }\right) \left\langle \phi _{%
\text{$fi$}}\right| \left( \hat{X}\hat{P}-\hat{P}\hat{X}\right) \left| \phi
_{\text{$fi$}}\right\rangle \\
&&+\frac{gt}{\hbar }\func{Im}\left( \frac{\left\langle F\right| \hat{A}%
\left| I\right\rangle }{\left\langle F|I\right\rangle }\right) \left\langle
\phi _{\text{$fi$}}\right| \left( \hat{X}\hat{P}+\hat{P}\hat{X}\right)
\left| \phi _{\text{$fi$}}\right\rangle \\
&=&gt\func{Re}\left( \frac{\left\langle F\right| \hat{A}\left|
I\right\rangle }{\left\langle F|I\right\rangle }\right) .
\end{eqnarray}%
Here, $\left\langle {}\right\rangle _{\text{$fi$}}$ is used to signify the
expectation value of a pointer observable only in the subensemble of
measured systems that start in state $\left| I\right\rangle $ and are later
post-selected in the state $\left| F\right\rangle .$ \ Similarly, the
momentum expectation value is given by%
\begin{eqnarray}
\left\langle \hat{P}\right\rangle _{\text{$fi$}} &=&\left\langle \phi _{%
\text{$fi$}}\right| \hat{P}\left| \phi _{\text{$fi$}}\right\rangle =\frac{%
-igt}{\hbar }\func{Re}\left( \frac{\left\langle F\right| \hat{A}\left|
I\right\rangle }{\left\langle F|I\right\rangle }\right) \left\langle \phi _{%
\text{$fi$}}\right| \left( \hat{P}^{2}-\hat{P}^{2}\right) \left| \phi _{%
\text{$fi$}}\right\rangle \\
&&+\frac{gt}{\hbar }\func{Im}\left( \frac{\left\langle F\right| \hat{A}%
\left| I\right\rangle }{\left\langle F|I\right\rangle }\right) \left\langle
\phi _{\text{$fi$}}\right| \left( \hat{P}^{2}+\hat{P}^{2}\right) \left| \phi
_{\text{$fi$}}\right\rangle \\
&=&\frac{\hbar gt}{2\sigma ^{2}}\func{Im}\left( \frac{\left\langle F\right| 
\hat{A}\left| I\right\rangle }{\left\langle F|I\right\rangle }\right) .
\end{eqnarray}

The shifts from zero in both the $\hat{X}$ and $\hat{P}$ expectation values
are proportional to the real and imaginary parts, respectively, of the weak
value $\left\langle \hat{A}\right\rangle _{W}$ which is defined as%
\begin{equation}
\left\langle \hat{A}\right\rangle _{W}\equiv \frac{\left\langle F\right| 
\hat{A}\left| I\right\rangle }{\left\langle F|I\right\rangle }.
\end{equation}%
In fact, AAV showed that for sufficiently weak coupling $\langle x|\phi
_{fi}\rangle $, the final pointer state, will be $\left( \sqrt{2\pi }\sigma
\right) ^{-\frac{1}{2}}\exp \left( -\left( x-\left\langle \hat{A}%
\right\rangle _{W}\right) ^{2}/4\sigma ^{2}\right) $, unchanged except for a
shift by the weak value.

It has been argued that it is the backaction of the measurement on the
measured system that leads to a finite $\func{Im}\left\langle \hat{A}%
\right\rangle _{W}$ and thus a nonzero $\left\langle \hat{P}\right\rangle _{%
\text{fi}}$ \cite{aeweak}. \ In addition, as the measurement becomes weaker $%
\left\langle \hat{P}\right\rangle _{\text{$fi$}}$ becomes more and more
difficult to determine; $\left\langle \hat{P}\right\rangle _{\text{$fi$}}$
decreases with $gt/\sigma ^{2}$ whereas the width $\Delta \hat{P}$ decreases
as $1/\sigma $. Some have gone as far as to define the weak value as $\func{%
Re}\left( \frac{\left\langle F\right| \hat{A}\left| I\right\rangle }{%
\left\langle F|I\right\rangle }\right) $ \cite{wisemanQED}. \ Nonetheless,
we will show that $\left\langle \hat{P}\right\rangle _{\text{$fi$}}$ should
not be interpreted as an insignificant artifact of the weak measurement
procedure and has an integral role in measuring the $N$th-order joint weak
value.

One can express the full weak value in terms of the two expectation values
of the pointer,%
\begin{eqnarray}
\left\langle \hat{A}\right\rangle _{W} &=&\func{Re}\left\langle \hat{A}%
\right\rangle _{W}+i\func{Im}\left\langle \hat{A}\right\rangle _{W} \\
&=&\frac{2\sigma }{gt}\left\langle \frac{1}{2\sigma }\hat{X}+i\frac{\sigma }{%
\hbar }\hat{P}\right\rangle _{\text{$fi$}}.  \label{simpleweak}
\end{eqnarray}%
In their derivation of \ weak values, AAV made the natural choice of a
Gaussian for the initial pointer state, as do we. \ This state also happens
to be the ground state $\left| 0\right\rangle $ of a harmonic oscillator
with mass $m$ and frequency $\omega $. \ For illustration, if one
reparameterizes the width of the Gaussian in terms of $m\omega $ such that $%
\sigma =\sqrt{\hbar /2m\omega }$ it becomes apparent that the operator in
the expectation value in Eq. (\ref{simpleweak}) is just the familiar
lowering operator, 
\begin{equation}
\hat{a}=\sqrt{\frac{m\omega }{2\hbar }}\hat{X}+i\sqrt{\frac{1}{2m\omega
\hbar }}\hat{P}.
\end{equation}%
The operator in Eq. (\ref{simpleweak}) will transform the pointer just as
the lowering operator does, even though the pointer is not actually in a
harmonic potential. \ This fact will simplify some of the following
calculations. \ Furthermore, now the weak value can be re-expressed as: 
\begin{equation}
\left\langle \hat{A}\right\rangle _{W}=\frac{2\sigma }{gt}\left\langle 
\hat{a}\right\rangle _{\text{$fi$}}.  \label{annihilation}
\end{equation}

To our knowledge, this is the first time in the literature that this simple
but important relationship between the annihilation operator and weak
measurement has been described. The reason the annihilation operator is
related to the weak value can be understood as follows. \ When the coupling
is sufficiently weak, the expansion in Eq. (\ref{expansion}) shows that the
largest pointer amplitude is left unchanged in the ground state. \ The
interaction Hamiltonian shifts some of the pointer state into the first
excited state by creating a small amplitude, proportional to $gt\hat{A}$,\
for the $\left| 1\right\rangle $ state. \ \ If we restrict ourselves to the
post-selected subensemble, as in Eq. (\ref{post select pointer}), then this
small amplitude changes to be proportional to $gt\left\langle \hat{A}%
\right\rangle _{W}$. \ The annihilation operator removes the part of the
state that is left unchanged by the coupling,\ leaving only the shifted
component. \ In other words, the annihilation operator isolates only that
part of the pointer state that is changed by the interaction.

We now move on to a derivation of $N$th-order joint weak values. \ In this
section, we combine the strategy outlined in the introduction for measuring
joint observables with the use of the annihilation operator to extract the
weak value. \ As in previous works, to measure the operator $\hat{M}%
=\prod_{j=1}^{N}\hat{A}_{j}$ we apply $N$ separate von Neumann interactions
coupling each $\hat{A}_{j}$ to its own pointer, as in Eq. (\ref{nvn}) \cite%
{twoweak,nweak}. \ To simplify the expressions to come we set all $g_{j}$ to
be equal and rewrite the momentum operators $\hat{P}_{j}$ in terms of the
respective raising and lowering operators, $\hat{a}_{j}^{\dagger }$ and $%
\hat{a}_{j}$, for each of the pointers, 
\begin{equation}
\mathcal{H}=i\frac{\hbar g}{2\sigma }\sum_{j=1}^{N}\hat{A}_{j}\left( \hat{a}%
_{j}^{\dagger }-\hat{a}_{j}\right) .
\end{equation}%
Now we require $N$ different pointers, all beginning in an initial state
defined by Eq. (\ref{pointer}). \ The total initial pointer state can be
described by the ground state of $N$ harmonic oscillators: 
\begin{equation}
\left| \Phi \right\rangle =\prod_{j=1}^{N}\left| \phi _{j}\right\rangle
=\left| 0\right\rangle ^{\otimes N}.  \label{N pointer state}
\end{equation}%
Continuing, using the number-state notation to describe the pointer, we
calculate the state of the combined system after the interaction Hamiltonian
is applied,%
\begin{eqnarray}
\left| \Phi \right\rangle \left| I\right\rangle  &\rightarrow &\exp \left( 
\frac{-i\mathcal{H}t}{\hbar }\right) \left| 0\right\rangle ^{\otimes
N}\left| I\right\rangle =\left( 1-\frac{i\mathcal{H}t}{\hbar }+...\right)
\left| 0\right\rangle ^{\otimes N}\left| I\right\rangle   \label{first order}
\\
&=&\left( 1+\frac{gt}{2\sigma }\sum_{j=1}^{N}\hat{A}_{j}\left( \hat{a}%
_{j}^{\dagger }-\hat{a}_{j}\right) +...\right) \left| 0\right\rangle
^{\otimes N}\left| I\right\rangle  \\
&=&\left| 0\right\rangle ^{\otimes N}\left| I\right\rangle +\frac{gt}{%
2\sigma }\sum_{j=1}^{N}\hat{A}_{j}\left| 1_{j}\right\rangle \left|
I\right\rangle +...,
\end{eqnarray}%
where $\left| 1_{j}\right\rangle $ is the state where the jth pointer is in
the first-excited state and all the other pointers are in the ground state
(e.g. $\left| 0_{1}1_{2}0_{3}...0_{N}\right\rangle $). \ Here, we have
expanded the state in powers of $gt$.\ \ Eq. (\ref{first order}) shows that
to first order, the interaction Hamiltonian coupling the measuring device to
the system can displace only one of the $N$ pointers at a time. Simultaneous
shifts of multiple pointers come from higher-order terms in the propagator.
\ We are particularly interested in the $N$th term in the expansion, 
\begin{equation}
\frac{1}{N!}\left( \frac{-i\mathcal{H}t}{\hbar }\right) ^{N}=\frac{1}{N!}%
\left( \frac{gt}{2\sigma }\sum_{j=1}^{N}\hat{A}_{j}\left( \hat{a}%
_{k}^{\dagger }-\hat{a}_{k}\right) \right) ^{N}.  \label{Nth term}
\end{equation}%
This term is the lowest-order one in the expansion which can simultaneously
transfer all $N$ pointers into the first excited state (e.g. $\left|
1_{1}1_{2}1_{3}...1_{N}\right\rangle $). \ This state, which we label as $%
\left| 1\right\rangle ^{\otimes N},$ is created when each term in the above
sum supplies one raising operator.\ \ The terms in the sum can contribute
the $N$ distinct raising operators in any order and so the portion of Eq. (%
\ref{Nth term}) that creates the $\left| 1\right\rangle ^{\otimes N}$ state
is equal to 
\begin{equation}
\frac{1}{N!}\frac{gt}{2\sigma }\mathcal{\wp }\left\{ \hat{A}_{k}\hat{a}%
_{k}^{\dagger }\right\} _{N},
\end{equation}%
where $\mathcal{\wp }\left\{ \hat{L}_{k}\right\} _{N}$ denotes the sum of
all $N!$ orderings of the set of \ $N$ operators $\left\{ \hat{L}%
_{k}\right\} $. \ Note that these different orderings are only distinct when
the operators do not commute. \ \ The remaining portions of Eq. (\ref{Nth
term}) create states where at least one pointer is left in the initial state
(e.g. $\left| 2_{1}0_{2}1_{3}...1_{N}\right\rangle $). \ Projecting onto $%
\left\langle F\right| $ completes the post-selection and leaves us with,%
\begin{eqnarray}
\left\langle F\right| \exp \left( \frac{-i\mathcal{H}t}{\hbar }\right)
\left| 0\right\rangle ^{\otimes N} &=&\left| 0\right\rangle ^{\otimes
N}\left\langle F|I\right\rangle +\frac{gt}{2\sigma }\sum_{j=1}^{N}\left%
\langle F\right| \hat{A}_{j}\left| I\right\rangle \left| 1_{j}\right\rangle
+... \\
&&+\left( \frac{gt}{2\sigma }\right) ^{N}\frac{1}{N!}\left\langle F\right| 
\mathcal{\wp }\left\{ \hat{A}_{k}\right\} _{N}\left| I\right\rangle \left|
1\right\rangle ^{\otimes N}+...
\end{eqnarray}%
\ We renormalize the resulting $N$-pointer state $\left| \Phi _{\text{fi}%
}\right\rangle $ and then truncate the amplitude of each term at the lowest
nonzero order in $gt,$ 
\begin{equation}
\left| \Phi _{\text{$fi$}}\right\rangle =\left| 0\right\rangle ^{\otimes N}+%
\frac{gt}{2\sigma }\sum_{j=1}^{N}\frac{\left\langle F\right| \hat{A}%
_{j}\left| I\right\rangle }{\left\langle F|I\right\rangle }\left|
1_{j}\right\rangle +...+\left( \frac{gt}{2\sigma }\right) ^{N}\frac{1}{N!}%
\frac{\left\langle F\right| \mathcal{\wp }\left\{ \hat{A}_{k}\right\}
_{N}\left| I\right\rangle \left| 1\right\rangle ^{\otimes N}}{\left\langle
F|I\right\rangle }+...  \label{final N pointer state}
\end{equation}%
\ This is equivalent to dividing by $\left\langle F|I\right\rangle $, the
renormalization constant in the limit of no coupling.\ In analogy with Eq. (%
\ref{annihilation}), we now wish to take the expectation value of the
product of the annihilation operators for all $N$ pointers,

\ 
\begin{equation}
\hat{O}\equiv \prod_{j=1}^{N}\hat{a}_{j}.
\end{equation}%
In Eq. (\ref{final N pointer state}), the $\left| 1\right\rangle ^{\otimes
N} $ state is the lowest order term that does not go to zero when acted on
by $\hat{O}$; this term becomes, \ 
\begin{equation}
\hat{O}\left| \Phi _{\text{$fi$}}\right\rangle =\left( \frac{gt}{2\sigma }%
\right) ^{N}\frac{1}{N!}\frac{\left\langle F\right| \mathcal{\wp }\left\{ 
\hat{A}_{j}\right\} _{N}\left| I\right\rangle }{\left\langle
F|I\right\rangle }\left| 0\right\rangle ^{\otimes N}+O\left( \left(
gt\right) ^{N+1}\right) .
\end{equation}%
Clearly, to lowest nonzero order the expectation value then becomes,%
\begin{eqnarray}
\left\langle \hat{O}\right\rangle _{\text{$fi$}} &=&\left\langle \Phi _{%
\text{$fi$}}\right| \hat{O}\left| \Phi _{\text{$fi$}}\right\rangle \\
&=&\left\langle 0\right| \left( \frac{gt}{2\sigma }\right) ^{N}\frac{1}{N!}%
\frac{\left\langle F\right| \mathcal{\wp }\left\{ \hat{A}_{j}\right\}
_{N}\left| I\right\rangle }{\left\langle F|I\right\rangle }\left|
0\right\rangle \\
&=&\left( \frac{gt}{2\sigma }\right) ^{N}\frac{1}{N!}\frac{\left\langle
F\right| \mathcal{\wp }\left\{ \hat{A}_{j}\right\} _{N}\left| I\right\rangle 
}{\left\langle F|I\right\rangle }.  \label{N expectation}
\end{eqnarray}%
The next lowest order term in the expectation value corresponds to any of
the $N$ pointers undergoing an extra pair of transitions (i.e., a pointer is
raised to $\left| 2\right\rangle $ and subsequently lowered back to $\left|
1\right\rangle $). \ Consequently it will be reduced in size by a factor of $%
2\left( \frac{gt}{2\sigma }\right) ^{2}$ compared to the lowest order term.
\ Using Eq. (\ref{N expectation}) the $N$th-order joint weak value can now
be expressed in the simple formula%
\begin{equation}
\frac{1}{N!}\left\langle \mathcal{\wp }\left\{ \hat{A}_{j}\right\}
_{N}\right\rangle _{W}=\left\langle \prod_{j=1}^{N}\hat{a}_{j}\right\rangle
_{\text{$fi$}}\left( \frac{2\sigma }{gt}\right) ^{N}.
\end{equation}

It is often the case that each operator $\hat{A}_{j}$ acts on a different
particle, ensuring that all $\hat{A}_{j}$ commute. This allows the further
simplification of the $N$th-order joint weak value to%
\begin{equation}
\left\langle \prod_{j=1}^{N}\hat{A}_{j}\right\rangle _{W}=\left\langle
\prod_{j=1}^{N}\hat{a}_{j}\right\rangle _{\text{$fi$}}\left( \frac{2\sigma }{%
gt}\right) ^{N}.  \label{the formula}
\end{equation}%
For commuting observables, the magnitude of the simultaneous shift in the $N$
pointers that results from concurrent kicks from all $N$ terms in the
Hamiltonian in Eq. (\ref{nvn}) is proportional to the shift in one pointer
created by a single von Neumann Hamiltonian for measuring operator $\hat{M}$%
. \ The role of the annihilation operators is to isolate this simultaneous
pointer shift from the total uncorrelated shifts of the $N$ pointers and
thus duplicate the action of $\mathcal{H}=g\hat{M}\hat{P}$, without the need
for multiparticle interactions.

Since Eq. (\ref{the formula}) requires the measurement of the annihilation
operator, which is not Hermitian, one might think the expression is, in
principle, unmeasurable. \ In fact, if one expands the annihilation operator
in terms of $\hat{X}$ and $\hat{P}$ for each pointer then one is simply left
with expectation values of products of $\hat{X}$ or $\hat{P}$ for each
pointer. \ One then measures $\hat{X}$ in one ensemble of pointers and $\hat{%
P}$ in an identically-prepared ensemble.

The expression in Eq. (\ref{the formula}) is the central result of this
work. \ As in previous papers, this result shows how one can practically
measure a joint weak value even without the multiparticle interactions the
AAV method requires \cite{twoweak,nweak}. \ However, this expression is much
more elegant and it makes it clear that the annihilation operator plays a
key role in joint weak measurements. \ Specifically, with the use of the
annihilation operator, the similarity to the strong measurement expectation
value in Eq. (\ref{strong}) is apparent. \ For strong measurement, the
equivalent expectation value to the $N$th-order joint weak value is%
\begin{equation}
\left\langle \prod_{j=1}^{N}\hat{A}_{j}\right\rangle =\left\langle
\prod_{j=1}^{N}\hat{X}_{j}\right\rangle \left( \frac{1}{gt}\right) ^{N}.
\end{equation}%
The similarity is striking and makes a good case for the use of the
annihilation operator in the understanding of weak values.

Lets compare Eq. (\ref{the formula}) to the previous results for the $N$%
th-order joint weak value \cite{nweak}. \ \ In the previous paper, it was
expressed recursively in terms of two $N$th-order correlations between the
pointers and to $N$ different joint weak values of order $N-1$. \ Utilizing
this recursive formula, the $N$th-order joint weak value can be expressed
purely in terms of the expectation value of position and momentum
correlations. \ This expression includes $2^{N+1}-2$ distinct correlations
of various orders, although most will be close to the $N/2$ order as the
number of distinct expectation values at each order follows the binomial
distribution. \ In comparison, Eq. (\ref{the formula}) relates the $N$%
th-order joint weak value to $2^{N}$ correlations in the positions and
momenta of all $N$ pointers and so requires roughly half the number of
expectation values as the final result from the previous paper (but of
higher order).

As a specific example of the use of Eq. (\ref{the formula}), the weak value
of the product of two spin components $S_{1x}S_{2y}$ would be,%
\begin{eqnarray}
\left\langle S_{1x}S_{2y}\right\rangle _{W} &=&\left( \frac{2\sigma }{gt}%
\right) ^{2}\left\langle \hat{a}_{1}\hat{a}_{2}\right\rangle _{\text{$fi$}}
\\
&=&\left( \frac{2\sigma }{gt}\right) ^{2}\left\langle \left( \frac{1}{%
2\sigma }\hat{X}_{1}+i\frac{\sigma }{\hbar }\hat{P}_{1}\right) \left( \frac{1%
}{2\sigma }\hat{X}_{2}+i\frac{\sigma }{\hbar }\hat{P}_{2}\right)
\right\rangle _{\text{$fi$}}.
\end{eqnarray}%
The real and imaginary parts of the weak value are then%
\begin{eqnarray}
\func{Re}\left\langle S_{1x}S_{2y}\right\rangle _{W} &=&\left( \frac{1}{gt}%
\right) ^{2}\left( \left\langle \hat{X}_{1}\hat{X}_{2}\right\rangle _{\text{$%
fi$}}-\frac{4\sigma ^{4}}{\hbar ^{2}}\left\langle \hat{P}_{1}\hat{P}%
_{2}\right\rangle _{\text{$fi$}}\right) \\
\func{Im}\left\langle S_{1x}S_{2y}\right\rangle _{W} &=&\frac{2\sigma ^{2}}{%
\hbar }\left( \frac{1}{gt}\right) ^{2}\left( \left\langle \hat{X}_{1}\hat{P}%
_{2}\right\rangle _{\text{$fi$}}+\left\langle \hat{P}_{1}\hat{X}%
_{2}\right\rangle _{\text{$fi$}}\right) .
\end{eqnarray}%
The importance of the pointer momentum shift is demonstrated in the above
example. \ With our measurement technique even the real part of weak value
is related to the pointers' momenta, $\hat{P}_{1}$ and $\hat{P}_{2}$ . \ \
In general, the momentum and position observables for each of the $N$
pointers will appear in the expression for the real part of the $N$th-order
joint weak value.

Note that like single weak measurements, this method for measuring the $N$%
th-order joint weak value is not limited to the particular interaction or
pointer used in our measurement model \cite{arbpointer}. \ For example, one
can perform a derivation very similar to the one presented here where a
spin, as opposed to position, pointer is used. \ For a spin pointer, the
Hamiltonian would be $\mathcal{H}=-g\hat{A}\hat{S}_{y}=ig\hat{A}(\hat{S}%
_{z}^{+}-\hat{S}_{z}^{-})/2$, where $\hat{S}_{i}^{+}$ and $\hat{S}_{i}^{-}$
are the raising and lowering operators for the $\hat{S}_{i}$ basis. \ The
initial pointer state would be the lowest eigenstate of $\hat{S}_{z},$ with
eigenvalue $-\hbar s$. \ \ In this case, the expression for the $N$th-order
joint weak value in terms of $N$ spin pointers is%
\begin{equation}
\left\langle \prod_{j=1}^{N}\hat{A}_{j}\right\rangle _{W}=\left\langle
\prod_{j=1}^{N}\hat{S}_{jz}^{-}\right\rangle _{\text{$fi$}}\left( \frac{1}{%
gt\hbar s}\right) ^{N},
\end{equation}%
where $\hat{S}_{jz}^{-}$ is the z-basis lowering operator for the jth
pointer and all $\hat{A}_{j}$ are assumed to commute. An important advantage
of using spin is the absence of unequal coefficients in the expression for
the lowering operator. \ This puts the shifts in the pointer observable and
its conjugate on equal footing. \ Using such a pointer means that the
physical shift in the conjugate observable does not become smaller as the
measurement becomes weaker. \ Expectation values are also particularly easy
to measure for spins (and polarizations), especially spin 1/2 systems since
there are only two basis states which need to be projected onto. \ For
instance, the $N$th-order joint weak value would only require $2^{2N}$
measurements in total if $N$ spin 1/2 pointers were used.

In the present work, we have greatly simplified a recent extension of weak
measurement which makes the experimental investigation of composite, or
joint observables possible \cite{twoweak,nweak}. \ We have shown that when
single and joint weak values are expressed as expectation values of
annihilation operators, they take on a surprisingly elegant form very
similar to that seen in standard strong measurement. \ This form is easily
generalized to any measurement device in which the initial pointer state is
the eigenstate of an appropriate lowering operator. \ With the extension,
the weak measurement of joint observables only requires the same apparatus
that one would need to weakly measure each of the component observables
separately. Joint observables are central to the detection and utilization
of entanglement in multiparticle systems. \ The weak measurement of these
observables should be particularly useful for investigating post-selected
systems such as those that have been used produce novel\ multiparticle
entangled states or those that implement quantum logic gates \cite%
{loqc,exploqc}.\ \ \ 

\section*{Acknowledgments}

This work was supported by ARC Seibersdorf Research GmbH, the Austrian
Science Foundation (FWF), project number SFB 015 P06, NSERC, and the
European Commission, contract number IST-2001-38864 (RAMBOQ). We would like
to thank Aephraim Steinberg and Morgan Mitchell for helpful discussions.

\end{document}